\documentclass[aps,prb,reprint,twocolumn,groupedaddress,amsmath,amssymb]{revtex4-1}

\usepackage[usenames,dvipsnames]{color}
\usepackage{graphicx,dcolumn,bm,amsmath,amssymb}
\usepackage{pslatex}

\begin{document}


\title{Influence of interstitial Mn on local structure and magnetism in Mn$_{1+\delta}$Sb}


\author{Joshua A. Kurzman}
\author{Andrew J. Martinolich}
\author{James R. Neilson}
\email[]{james.neilson@colostate.edu}
\affiliation{Department of Chemistry, Colorado State University, Fort Collins, Colorado 80523-1872, United States}


\date{\today}

\begin{abstract}
We report x-ray total scattering and pair distribution function (PDF) studies of the structural relaxation around interstitial manganese (Mn$_i$) in ferromagnetic Mn$_{1+\delta}$Sb ($0.03 \le \delta \le 0.23$) alloys, guided by density functional theory (DFT). Refinements to the experimental PDF using a crystallographically constrained structural model indicate an expansion in the equatorial plane of the Mn$_i$Sb$_5$ trigonal bipyramidal site, which introduces significant positional disorder in addition to the nominally-random occupation of interstitial voids. Observation of a weak diffuse signal near the symmetry-forbidden (001) reflection position is indicative of correlated disorder from the clustering of Mn$_i$.  Density functional relaxation of supercells approximating the $\delta = 0.08$, $0.15,$ and $0.23$ compositions provides improved models that accurately describe the short-range structural distortions captured in the PDFs. Such structural relaxation increases the DFT calculated moment on Mn$_i$, which aligns antiparallel to the primary Mn moments, but leads to insubstantial changes in the average Mn and Sb moments and moments of Mn and Sb proximal to interstitials, thus providing a more accurate description of the observed bulk magnetic properties.
\end{abstract}


\maketitle



\section{Introduction}

\begin{figure}[htb]
\centering \includegraphics[width=3.3in]{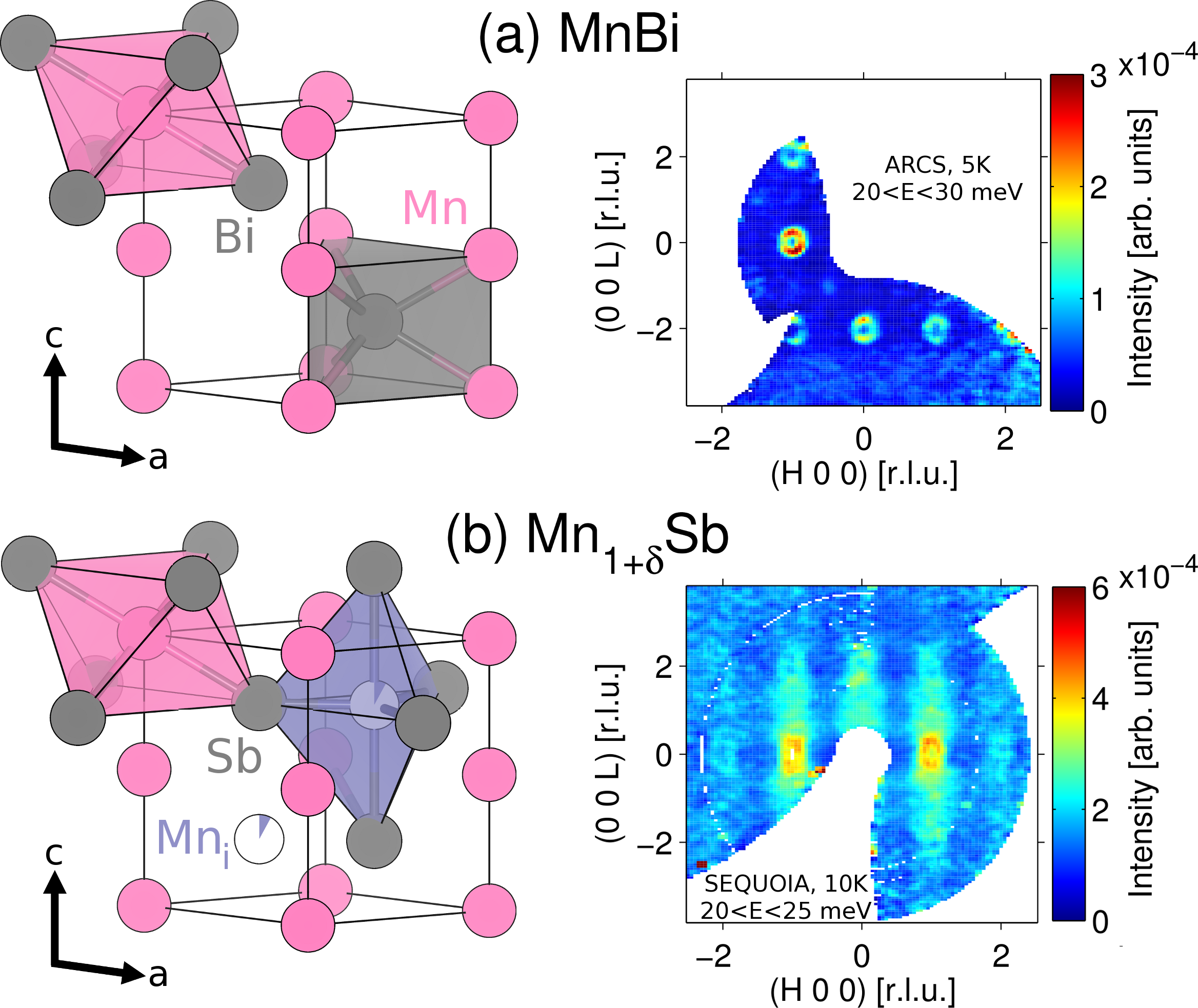}
\caption{(a) NiAs structure exemplified by the hard ferromagnetic phase MnBi [space group $P6_3/mmc$, with Mn on 2$a$ at (0,0,0) and Bi on 2$c$ at ($\frac{1}{3},\frac{2}{3},\frac{1}{4}$)], and inelastic neutron scattering data collected at 4\,K from a single crystal of MnBi. (b) Structure of Mn$_{1+\delta}$Sb highlighting the fractionally occupied, trigonal bipyramidal coordination environment of interstitial Mn [Mn$_i$, on 2$d$ site at ($\frac{1}{3},\frac{2}{3},\frac{3}{4}$)]. Note that Mn$_i$ is also in trigonal prismatic coordination with respect to the primary (fully occupied) Mn$_{\mathrm{Mn}}$ site.  Inelastic neutron scattering data collected at 10\,K on a single crystal of Mn$_{1.13}$Sb shows a pronounced diffuse component that is not observed for MnBi. Inelastic scattering figures adapted from Taylor \textit{et al.}, reference \onlinecite{Taylor:2015}.}
\label{fig:structure}
\end{figure}

Manganese pnictides of the hexagonal NiAs structure type (\textit{e.g.}\ MnAs, Mn$_{1+\delta}$Sb, MnBi; Figure \ref{fig:structure}) display a wealth of functional magnetic properties. The class of materials includes potential candidates for rare earth-free permanent magnets,\cite{HEIKES:1955,Zarkevich:2014}
 magneto-optical data storage,\cite{WILLIAMS:1957,CHEN:1968,UNGER:1972,BAI:1984}
  and magnetic refrigeration.\cite{Wada:2001,Sun:2008}
Manganese antimonide is known to only exist as a manganese-rich phase, Mn$_{1+\delta}$Sb.\cite{TERAMOTO:1968,VANYARKHO:1988}
The wide compositional region shown on the equilibrium phase diagram extends from $0.02 \lesssim \delta \lesssim 0.23$,\cite{Okamoto-H.:2006}
 with excess Mn occupying the interstitial site (Figure \ref{fig:structure}b) in trigonal bipyramidal coordination by Sb.
 Increasing amounts of interstitial Mn (Mn$_i$) lengthen the $a$ and shorten the $c$ lattice parameters, which accompanies reductions of the saturation magnetization, spin-reorientation temperature ($T_{\mathrm{SR}}$),\cite{Reimers:1982,Markandeyulu:1987,Taylor:2015}
  and the Curie temperature ($T_{\mathrm{C}}$).\cite{TERAMOTO:1968,OKITA:1968}
 Magnetic ordering of the primary Mn sites occurs parallel to $c$ at high temperature, but it changes to the $ab$ plane below $T_{\mathrm{SR}}$.\cite{TAKEI:1963}
 
 Although the bulk magnetic properties are well known, the microscopic nature of the magnetism associated with Mn$_i$ has been equivocal.
A straightforward explanation of the reduced magnetization in more Mn-rich samples is that Mn$_i$ is aligned antiparallel to the ferromagnetic Mn$_{\mathrm{Mn}}$ atoms, that is, Mn$_{1+\delta}$Sb is ferrimagnetic. (Here we use Kr\"oger--Vink notation to differentiate interstitial Mn$_i$ and nominal Mn$_{\mathrm{Mn}}$ atoms.) Nonpolarized neutron diffraction studies are consistent with this expectation.\cite{TAKEI:1963,OKITA:1968,Taylor:2015}
 However, polarized neutron scattering analyses have suggested that Mn$_i$ has no moment. \cite{YAMAGUCHI:1976,YAMAGUCHI:1980,Watanabe:1980,REIMERS:1983} The reduction of magnetization is then explained on the basis of greater orbital overlap between Mn atoms upon reduction of the $c$ lattice constant,
  and to a local perturbation of the Mn moment when in proximity to Mn$_i$.\cite{YAMAGUCHI:1976,Watanabe:1980,YAMAGUCHI:1980,REIMERS:1980,REIMERS:1983}
 
The question of whether Mn$_i$ carries a moment was recently re-addressed by Taylor \textit{et al.} via a combination of nonpolarized elastic and inelastic neutron scattering on an Mn$_{1.13}$Sb single crystal.\cite{Taylor:2015}
A critical feature of the modeling strategies adopted in previous \textit{polarized} neutron studies was the use of a highly aspherical magnetic form factor for Mn.\cite{Watanabe:1980,REIMERS:1983} In the recent work of Taylor and co-workers it was shown that such a model provides an inferior description of the  magnetic reflections -- particularly at low $Q$ where magnetic scattering is strongest -- relative to a model employing a spherical magnetic form factor and antiferromagnetic coupling between Mn$_i$ and Mn$_{\mathrm{Mn}}$.\cite{Taylor:2015} Density functional calculations enlisting small supercells further supported Mn$_{1+\delta}$Sb as a ferrimagnet. Additionally, a pronounced and gapless diffuse magnetic component was observed across all temperature regimes in the inelastic spectrum (Figure \ref{fig:structure}b), centered at (001). Similar or related features have also been noted by other groups.\cite{TAKEI:1963,Radhakrishna:1996} The location of the diffuse component is intriguing because the (001) reflection should be systematically absent from the structure factor; there is no diffuse scattering observed for stoichiometric (interstitial-free) MnBi (Figure \ref{fig:structure}a).
 Two plausible explanations were suggested: the diffuse scattering could arise from ($i$) correlated structural or magnetic disorder, or ($ii$) modification of the neighboring Mn$_{\mathrm{Mn}}$ moments by Mn$_i$.\cite{Taylor:2015}

An essential feature of the Mn$_{1+\delta}$Sb system that has been largely unaccounted for, both experimentally and computationally, is the local structure around Mn$_i$. It has long been recognized that the equatorial Mn$_i$--Sb contacts of the crystallographic Mn$_i$ site in the Mn$_{1+\delta}$Sb unit cell are quite short.\cite{YAMAGUCHI:1980,COEHOORN:1985}
 However, few studies have attempted to model the presumed structural relaxation.\cite{YAMAGUCHI:1980,Taylor:2015} Yamaguchi and Watanabe approximated that Mn$_i$ would be accommodated by distortions of only the Sb positions, the magnitude of the local displacements being derived from the concentration dependence of the lattice parameters.\cite{YAMAGUCHI:1980} This assumption was used in their analysis of polarized neutron diffuse scattering data. Models with no Mn$_i$ moment, but containing reduced moments on the first or second nearest-neighbor Mn atoms in proximity to Mn$_i$, were compared against calculated scattering for a model with Mn$_i$ aligned antiparallel. Although the best agreement was obtained for a reduction of the six nearest-neighbor Mn moments, an aspherical Mn magnetic form factor appears to have contributed to the result.\cite{Watanabe:1980,YAMAGUCHI:1980}

The utility of density functional modeling in describing local structural effects and defect physics is well demonstrated. A variety of approaches can be used to simulate defects (substitutional, interstitial, antisite, vacancy) and random alloying, the validity of which is reflected in agreement with a variety of experimental probes of local bonding arrangements, including nuclear magnetic resonance (NMR),\cite{Blanc:2011,Buannic:2012,Dervisoglu:2014}
 extended x-ray absorption fine structure (EXAFS),\cite{Grinberg:2002,Grinberg:2004a,Grinberg:2004,Levin:2011}
 and pair distribution function (PDF) data obtained from total scattering. \cite{Li:2007,Page:2007,White:2010,White:2010a,Kalland:2013}
When the concentration of defects is low, an effective strategy is to model a defect in a sufficiently large supercell such that the defect is approximately isolated (\textit{i.e.}, limiting the interaction between defects when periodic boundary conditions are applied).\cite{Van-de-Walle:2004}
 At higher levels of substitution, the creation of models on the basis of chemically reasonable, ordered supercells can be an elegant strategy to assess plausible bonding motifs.\cite{Li:2007,Page:2007}
Random alloys are inherently challenging systems to model in the framework of density functional theory (DFT), since even large supercells will necessarily contain elements of periodicity that are absent in a truly random structure, but strategies such as stochastic mixing and the special quasirandom structures approach often provide adequate approximations.\cite{Zunger:1990,WEI:1990,Levin:2011,Voas:2014} Mn$_{1+\delta}$Sb can be regarded as a type of random alloy, with a distribution of interstitials and vacancies over the crystallographic Mn$_i$ site.

In the present contribution, we examine local structural relaxation in Mn$_{1+\delta}$Sb using synchrotron x-ray total scattering and PDF analysis coupled with density functional modeling. The use of x rays rather than neutrons allows us to base our analysis on the local nuclear structure without interference from magnetic scattering. Particular attention is devoted to the Mn$_i$ environment and its calculated magnetic moment, as well as the compositional and structural influence on the moment of ferromagnetically aligned Mn$_{\mathrm{Mn}}$.
Structural relaxation has a pronounced effect on the Mn$_i$ environment and enhances electronic localization by a lengthening of equatorial Sb contacts, but it does not appear to influence the moments on Mn$_{\mathrm{Mn}}$ or Sb. A weak diffuse signal is observed near the symmetry-forbidden (001) Bragg position, suggestive of correlated disorder associated with clustering of Mn$_i$. These results are discussed in light of the recent work of Taylor {\textit et al.}\cite{Taylor:2015}

\section{Materials and Methods}

\subsection{Preparation of Mn$_{1+\delta}$Sb}
The series of polycrystalline Mn$_{1+\delta}$Sb ($0.01 \le \delta \le$ 0.23) samples were prepared in alumina crucibles from the elements (Mn flake, NOAH Technologies Corporation, 99.99\,\%; Sb shot, Alfa Aesar, 99\,\%). Antimony was recrystallized before use, and surface oxide was removed from Mn by heating overnight in an evacuated fused silica tube at 980\,$^\circ$C.  Appropriate stoichiometric mixtures sealed in evacuated silica tubes were melted at 930\,$^\circ$C for 16\,h and then quenched in water.  The obtained ingots were finely powdered in an agate mortar and pressed into pellets, then annealed under vacuum at 700\,$^\circ$C for 48\,h and water-quenched.

\subsection{Characterization}
Magnetization measurements were performed with a vibrating sample magnetometer on a Quantum Design, Inc. Dynacool PPMS. Room-temperature x-ray powder diffraction and x-ray total scattering measurements were conducted at the Advanced Photon Source of Argonne National Laboratory.

High-resolution synchrotron x-ray powder diffraction patterns were collected at the 11-BM-B beamline using an x-ray energy of about 30\,keV ($\lambda \approx 0.414$\,\AA).\cite{Lee:2008,Wang:2008,Toby:2009} Samples were diluted with amorphous SiO$_2$ to reduce the effects of absorption. Rietveld analyses were performed within the {\sc gsas/expgui} suite.\cite{Larson:2000cr,Toby:2001nx}

X-ray total scattering measurements were performed at the 11-ID-B beamline. Data were collected with an amorphous silicon area detector\cite{Chupas:2007}
using two x-ray energies, $\sim58$\,keV ($\lambda = 0.2114$\,\AA) and $\sim86$\,keV ($\lambda = 0.1430$\,\AA), at sample-to-detector distances of approximately 17\,cm and 19\,cm, respectively. Calibrations were performed by measurement of a CeO$_2$ standard at each condition. The electron density pair distribution function, $G(r)$, was obtained from background-subtracted scattering data by the \textit{ad-hoc} approach applied in PDFgetX3.\cite{Juhas:2013}
 The reduced scattering structure function, $S(Q)$, was transformed to $G(r)$ using a maximum momentum transfer of $Q_{\mathrm{max}}=24$\,\AA$^{-1}$ for data collected at 58\,keV, and $Q_{\mathrm{max}} = 28$\,\AA$^{-1}$ for data collected at 86\,keV. A powdered nickel standard was used to determine the resolution truncation parameters $Q_{\mathrm{damp}}$ and $Q_{\mathrm{broad}}$ used in PDFgui refinements.\cite{Farrow:2007} 
 Reverse Monte Carlo (RMC) simulations were conducted using the RMCprofile software,\cite{Tucker:2007} with models constrained to fit $S(Q)$ and $G(r)$ to capture both long- and short-range order. A ``closest approach'' constraint of 2.2\,\AA\ was applied to all atomic species to prevent the fitting of termination ripples in the low $r$ region of the PDF. Simulated diffraction patterns were computed using the DIFFaX software;\cite{TREACY:1991} input files are provided in the supplemental material.

\subsection{Computational Details}
Density functional theory (DFT) calculations were performed using the Vienna \textit{ab-initio} simulation package (VASP),\cite{Kresse:1994fk}
with interactions between the cores (Mn:[Ar], Sb:[Kr]4d$^{10}$) and valence electrons described using the projector augmented wave (PAW) method.\cite{Kresse:1999} Calculations were performed within the generalized gradient approximation (GGA) using the functional of Perdew, Burke, and Ernzerhof (PBE)\cite{Perdew:1996kx} to account for the effects of exchange and correlation. An energy cutoff of 350\,eV and a $\Gamma$-centered $4\times4\times4$ $k$-point sampling were employed for all calculations.
Since PDF analysis is a measurement of the ensemble average of all pairwise correlations in a sample, a series of 10 distinct initial supercell configurations were constructed for each of three compositions, Mn$_{48}$(Mn$_i$)$_{4}$Sb$_{48}$, Mn$_{48}$(Mn$_i$)$_{7}$Sb$_{48}$, and Mn$_{48}$(Mn$_i$)$_{11}$Sb$_{48}$ (where Mn$_i$ denotes interstitial manganese), which respectively provide compositions close to Mn$_{1.08}$Sb, Mn$_{1.15}$Sb, and Mn$_{1.23}$Sb. Parent Mn$_{48}$Sb$_{48}$ supercells were built from orthonormal $C$-centered orthorhombic supercells of the conventional crystallographic lattice, with dimensions reflecting the lattice constants obtained by Rietveld refinement. Interstitial Mn positions were selected stochastically with the aid of RMCprofile\cite{Tucker:2007}
 by filling all of the interstitial positions with Mn or vacancy (dummy) atoms and distributing Mn$_i$ by performing short simulations in which Mn$_i$ atoms were ``swapped'' with vacancy atoms; simulations lasting 6 seconds typically produced 50,000 atom swap moves, leading to a robust ensemble of starting structural configurations. Relaxations of the supercells were performed at fixed cell dimensions, allowing only the atomic positions to change, and they were deemed to have converged when the forces on all ions were less than 0.01\,eV\,\AA$^{-1}$. All calculations were performed with spin polarization, using starting spin configurations in which Mn$_i$ was aligned antiparallel to Mn. For select configurations, test calculations were performed with an on-site Coulomb repulsion term on Mn (GGA+$U$), or with spin--orbit coupling; neither the inclusion of a Hubbard $U$ nor the effects of the spin--orbit interaction produced any qualitative changes to the results discussed below.

\section{Results and discussion}

\subsection{Crystallographic analyses and bulk magnetic properties}

 \begin{figure}[!ht]
\centering \includegraphics[width=3in]{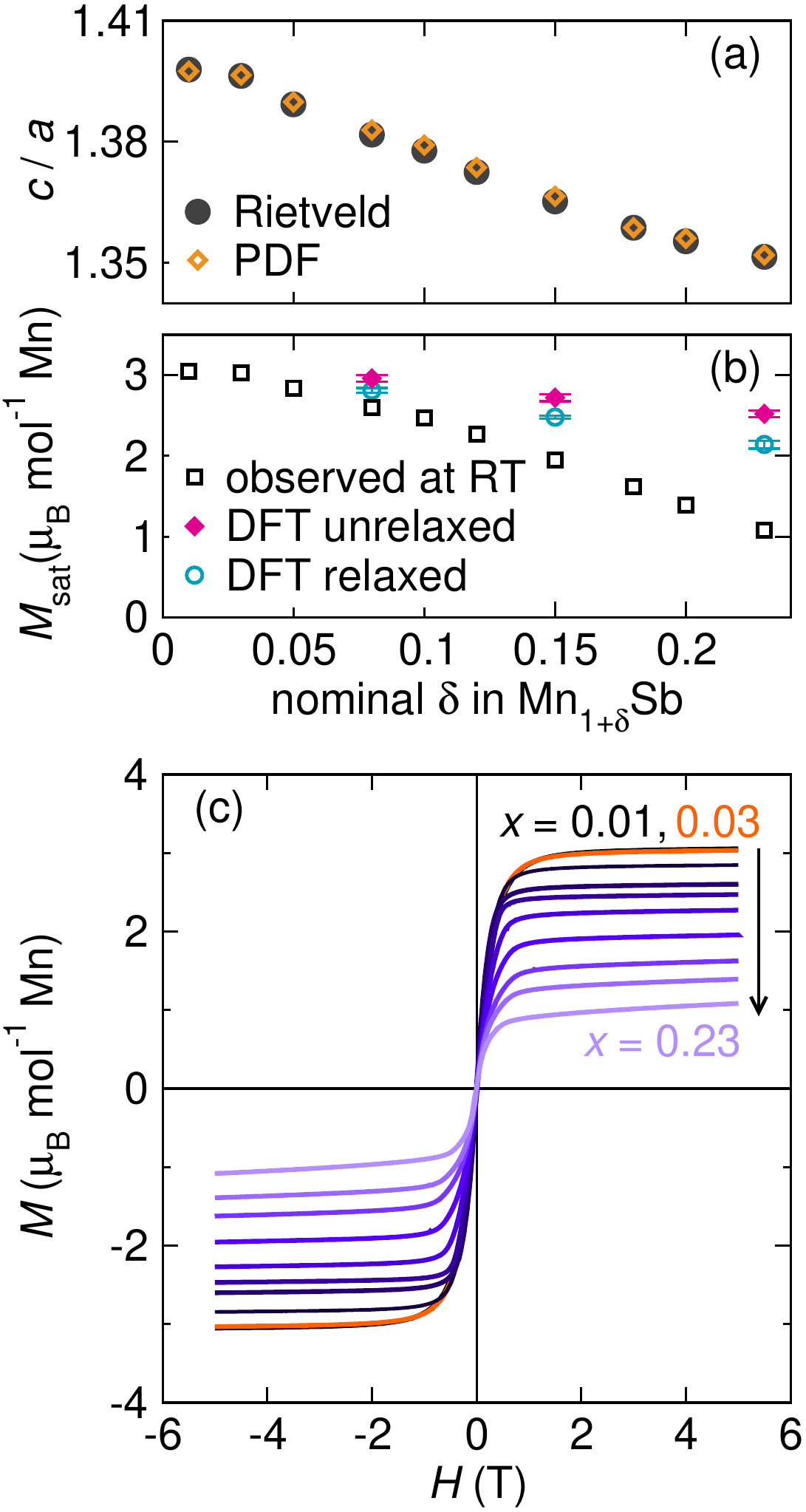}
\caption{(a) Axial ($c$/$a$) ratio of the Mn$_{1+\delta}$Sb lattice constants as a function of the nominal interstitial Mn content, $\delta$, showing V\'{e}gard law behavior over most of the compositional range and excellent agreement between Rietveld and PDF refinements and literature values.\cite{TERAMOTO:1968} Error bars are smaller than the symbols.  (b) Room temperature saturation magnetization as a function of $\delta$, and bulk magnetization at 0\,K obtained from DFT calculations of unrelaxed and relaxed supercells of Mn$_{48}$(Mn$_i$)$_x$Sb$_{48}$ ($x=4, 7, 11$); DFT moments represent the average of 10 distinct configurations at each composition, with error bars representing the standard deviation of the average. (c) Room temperature magnetization hysteresis curves of the Mn$_{1+\delta}$Sb series of samples. Note that the magnetization curves for $\delta=0.01$ and $\delta=0.03$ nearly overlay, which is consistent with the samples' $c/a$ ratios being nearly the same.}
\label{fig:average}
\end{figure}

\begin{figure}[!ht]
\centering \includegraphics[width=3.3in]{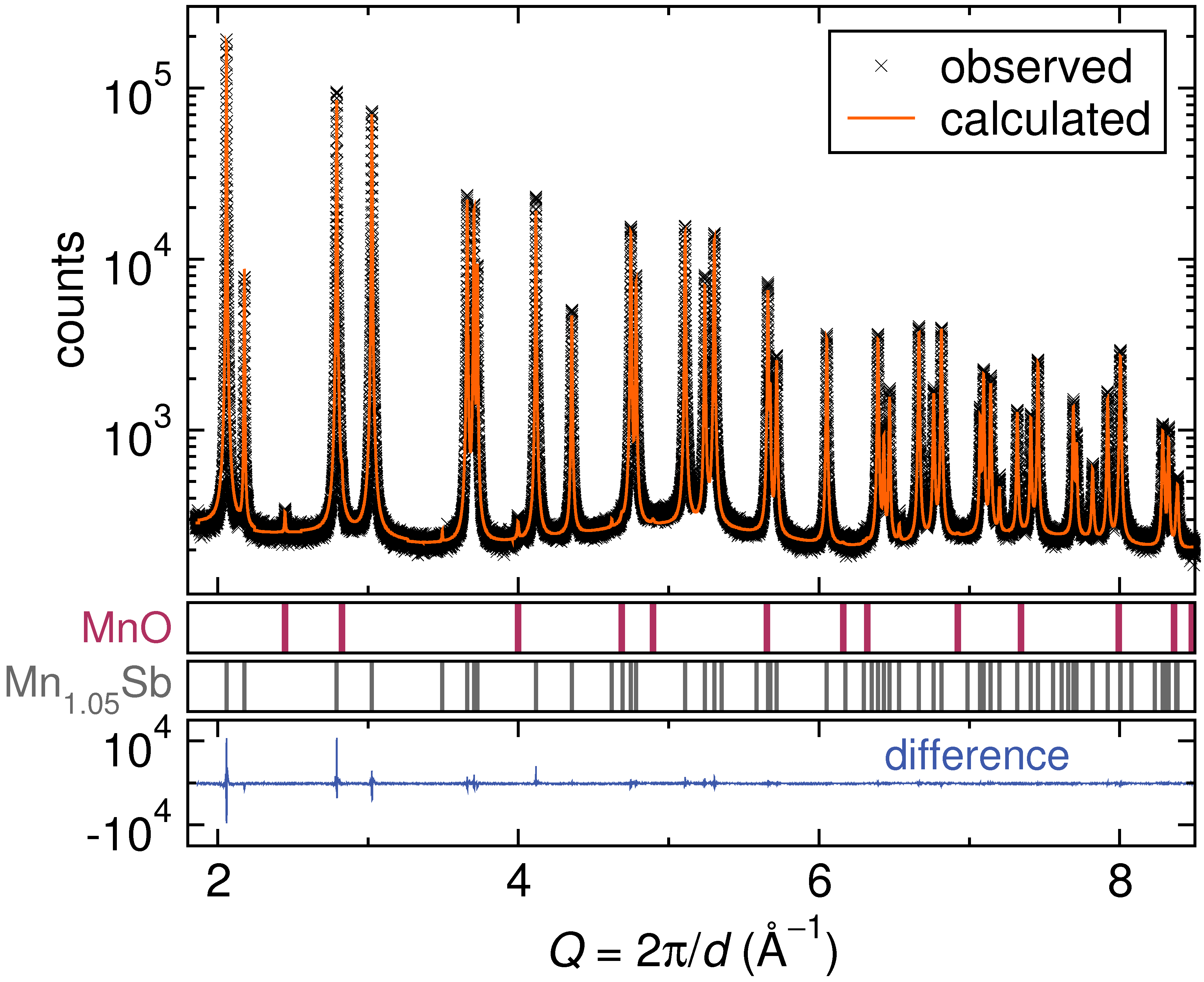}
\caption{Representative Rietveld refinement of Mn$_{1.05}$Sb against high resolution synchrotron x-ray diffraction data (11-BM, Advanced Photon Source). To aid in visual inspection of the fit, the low and high $Q$ regions of the diffraction data are omitted. Reflection positions for Mn$_{1.05}$Sb and the MnO impurity are shown as solid vertical lines in the two middle panels.}
\label{fig:rietveld}
\end{figure}

The series of Mn$_{1+\delta}$Sb samples ($0.01 \le \delta \le 0.23$) span approximately the entire compositional range and display V\'{e}gard law behavior of the lattice parameters between $0.03 \le \delta \lesssim 0.23$, as shown by the linear trend in $c/a$ ratio as a function of $\delta$, Figure \ref{fig:average}a. 
Small deviations from linearity are noted for the nominally $\delta=0.01$ and $\delta=0.23$ compositions, indicating proximity to the phase boundary extrema. The $\delta=0.01$ sample contains a small amount of antimony metal ($\le 1$\,wt.\,\%), and all of the samples contain small impurities of MnO, estimated by quantitative phase analysis to be $\lesssim1$\,wt.\,\%. There is no evidence of an Mn$_2$Sb impurity in the Mn$_{1.23}$Sb sample, suggesting the solubility limit had not been reached. Axial ratios obtained from Rietveld refinements of high-resolution synchrotron x-ray diffraction data (11-BM) agree very closely with the ratios obtained from real-space refinements of PDF data. A representative Rietveld refinement is presented in Figure \ref{fig:rietveld} for the $\delta=0.05$ sample, and a complete list of refined values for both Rietveld and PDF refinements is given in Table SM-1.  The Mn$_i$ occupancies estimated by Rietveld refinement agree well with the nominal stoichiometries across the $0.03 \leq \delta \leq 0.23$ range, with estimated errors in the site occupancy of 0.01.  Fractionally occupied Mn$_i$ has a site multiplicity of 2 (Wyckoff position 2$d$) in the crystallographic model, but this does not imply that interstitial atoms necessarily occur in pairs in a given unit cell of the  ``real'' material. Consider that for $\delta=0.05$ there is only one Mn$_i$ for every 10 unit cells [\textit{i.e.}\ Mn$_{20}$(Mn$_i$)$_1$Sb$_{20}$]: although the crystallographic cell indicates that Mn$_i$ populates both sites of the 2$d$ position {\em on average}, it is quite conceivable that interstitial atoms are present {\em locally} in just one of the two interstitial voids.

 Room-temperature saturation magnetizations (Figure \ref{fig:average}b), taken from the room-temperature hysteresis curves shown in Figure \ref{fig:average}c, decrease linearly as the quantity of interstitial Mn increases. Magnetization curves of the $\delta=0.01$ and $\delta=0.03$ samples are nearly overlaid, which is consistent with their comparable $c/a$ ratios and $\delta$ values obtained by Rietveld refinement.

\begin{figure}[!ht]
\centering \includegraphics[width=3.3in]{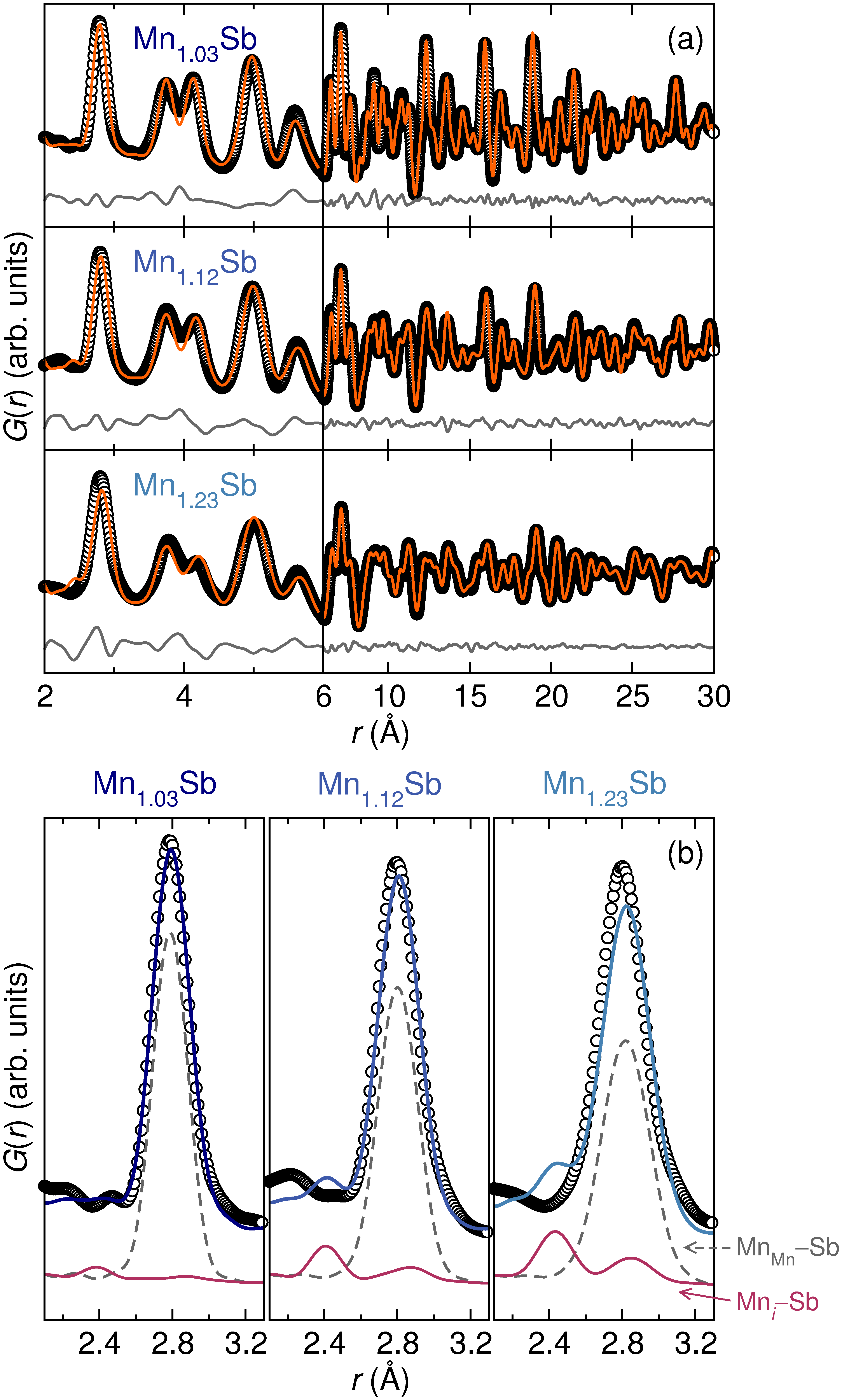}
\caption{PDF refinements (solid lines) against data (open circles) for samples of nominal compositions Mn$_{1.03}$Sb, Mn$_{1.12}$Sb, and Mn$_{1.23}$Sb, employing the crystallographic structural model in which interstitial Mn (Mn$_i$) is located on the $2d$ site and partially occupied. (a) Full-range fits, from 2 to 30\,\AA. The vertical line at $r=6$\,\AA\ demarcates different $x$-axis scales at shorter and longer $r$.  (b) Panels showing the first-nearest-neighbor region of the three full-range fits; select PDF partial contributions for Mn$_{\mathrm{Mn}}$--Sb (dotted line) and Mn$_i$--Sb (solid line) correlations are vertically offset for clarity.  As the amount of Mn$_i$ increases, the crystallographic structural model's deficiency becomes more apparent: there are shorter pair correlations present in the model than observed in the data, and these arise from equatorial Mn$_i$--Sb distances.
}
\label{fig:pdf}
\end{figure}

In general, PDFs of the Mn$_{1+\delta}$Sb series are well described by the conventional crystallographic cell (constrained by space-group symmetry), particularly at low concentrations of Mn$_i$ where there is limited sensitivity to its pair correlations. Real-space refinements (PDFgui) over the 2--30\,\AA\ range for Mn$_{1.03}$Sb, Mn$_{1.12}$Sb, and Mn$_{1.23}$Sb are shown in Figure \ref{fig:pdf}a, and they are representative of fit quality across the complete series of samples (Figure SM-1). These fits worsen as a function of increasing $\delta$ in the low $r$ region of the PDFs, but beyond about 5\,\AA\ the crystallographic structures provide very satisfactory fits to the data. In fact, for samples containing up to 10\,\% Mn$_i$, better statistical agreement is consistently obtained in refinements over the 6 to 30\,\AA\ range with a Mn$_{1.00}$Sb model (\textit{i.e.}, without Mn$_i$) than for Mn$_{1+\delta}$Sb models containing partially occupied Mn$_i$.

The corresponding first nearest-neighbor peaks, comprising contributions from Mn$_{\mathrm{Mn}}$--Sb, Mn$_i$--Sb and Mn$_i$--Mn$_{\mathrm{Mn}}$, are highlighted in Figure \ref{fig:pdf}b; the PDF partials for Mn$_{\mathrm{Mn}}$--Sb and Mn$_i$--Sb pair correlations are shown below in dotted and solid lines, respectively. For clarity, Mn$_{\mathrm{Mn}}$--Mn$_i$ partials are not shown as these contribute very weakly at the same distances as Mn$_{\mathrm{Mn}}$--Sb. The crystallographic model predicts contributions at approximately 2.4\,\AA, although none are observed experimentally; this intensity is derived from equatorial Mn$_i$--Sb correlations of the Mn$_i$ trigonal bipyramid. 
 The absence of observed intensity at this distance in $G(r)$ is evidence that structural relaxation occurs either by or around interstitial atoms, which is also reflected in the poorer description of peak intensity and breadth for the more Mn-rich samples. This result is quantitatively consistent for the PDFs obtained from total scattering with two different x-ray wavelengths. This feature is also robust considering the Nyquist sampling frequency dictated by $Q_{\mathrm{max}}$ (Figure SM-2).\cite{Farrow:2011}
  Comparing the crystallographic structural model against the observed data indicates that the equatorial Mn$_i$--Sb contacts are longer in the ``real'' material than reflected by the crystallographic cell. Qualitatively, a bond-lengthening around Mn$_i$ would be expected to increase electron localization and its magnetic moment.\cite{Anderson:1961}
 The ability to directly inspect PDF data and infer plausible structure--property relations is an indisputable asset of the technique. However, a more rigorous modeling approach is required to support and reconcile the impact of structural relaxation on the properties of Mn$_{1+\delta}$Sb.

\subsection{DFT: unifying local structure and magnetic properties}

\begin{figure}[!ht]
\centering \includegraphics[width=1.75in]{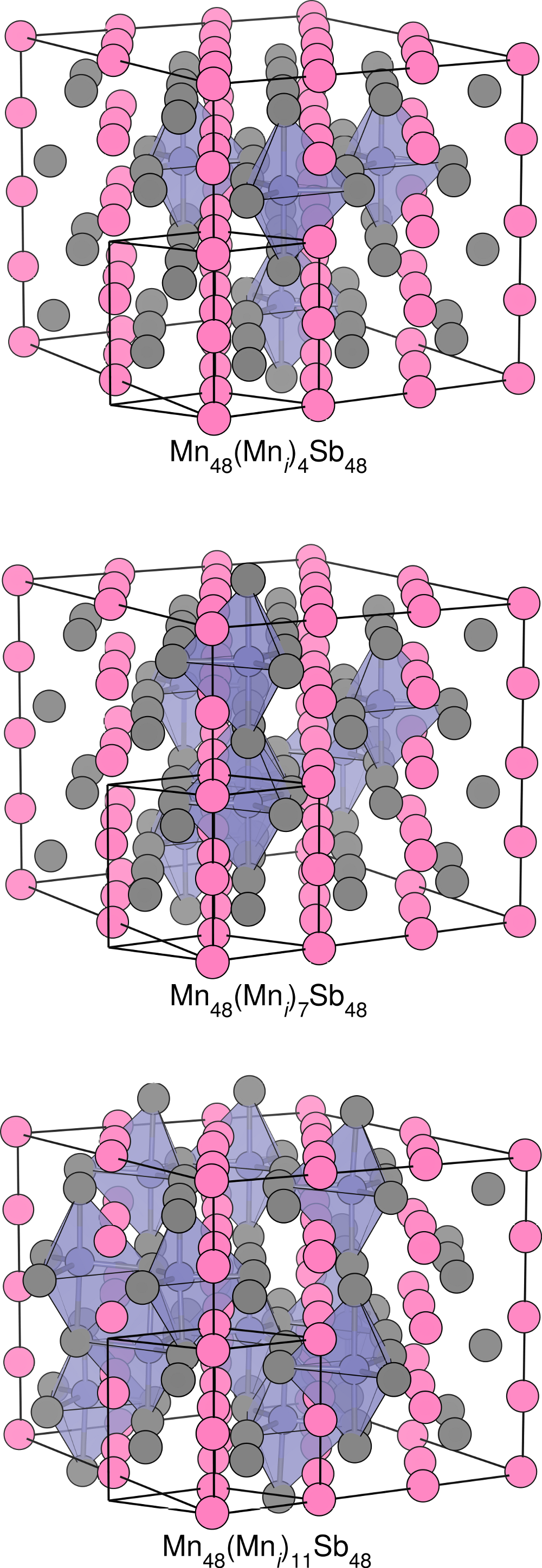}
\caption{Structural depictions of representative relaxed supercell configurations for Mn$_{48}$(Mn$_i$)$_{4}$Sb$_{48}$ (top), Mn$_{48}$(Mn$_i$)$_{7}$Sb$_{48}$ (middle), and Mn$_{48}$(Mn$_i$)$_{11}$Sb$_{48}$ (bottom). Mn$_{\mathrm{Mn}}$ is shown as light (pink) spheres, Sb as dark (gray) spheres, and Mn$_i$ (blue) in polyhedral rendering. An outline of the  conventional unit cell is shown for reference.}
\label{fig:supercells}
\end{figure}

A series of density functional calculations were performed to assess relaxation by and around interstitial atoms in Mn$_{1+\delta}$Sb.  To account for the disorder inherent to this system, a collection of 10 distinct supercell configurations was created for each of three compositions selected from the compositional range. All configurations are based on 96 atom MnSb supercells, with Mn$_{48}$(Mn$_i$)$_{4}$Sb$_{48}$, Mn$_{48}$(Mn$_i$)$_{7}$Sb$_{48}$, and Mn$_{48}$(Mn$_i$)$_{11}$Sb$_{48}$ formulas chosen to provide compositions close to Mn$_{1.08}$Sb, Mn$_{1.15}$Sb, and Mn$_{1.23}$Sb. The positions of Mn$_i$ in the supercell configurations were selected stochastically. Representative structural depictions of one configuration for each composition are shown in Figure \ref{fig:supercells}.
 To facilitate comparison with the PDF data, supercells for a given composition were constrained to dimensions commensurate with the experimentally determined unit-cell parameters.
 
\begin{figure}[!ht]
\centering \includegraphics[width=3in]{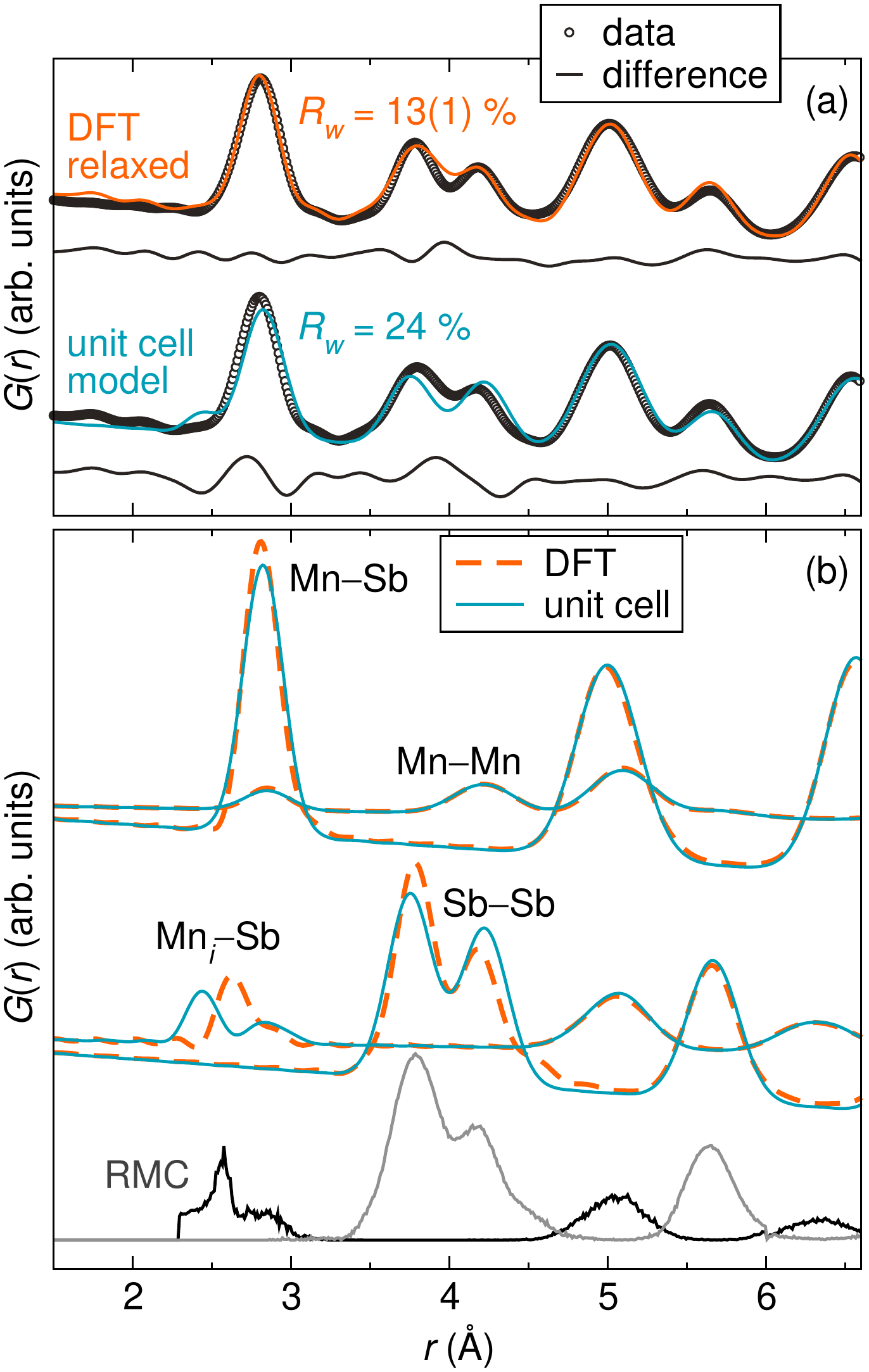}
\caption{(a) Comparison of PDF refinements for a crystallographically-constrained (unit cell) structural model and representative DFT-relaxed configuration against data collected for the sample of nominal composition Mn$_{1.23}$Sb. Refinement $R$-factors shown from fits in the range of 2 to 6\,\AA; $R_w$ for the DFT relaxed configuration represents the average for all 10 configurations, given with the standard deviation of the average. (b) Select atom-pair partials extracted from the DFT-relaxed and unit cell models, along with PDF partials generated from a reverse Monte Carlo simulation using a 9,990 atom supercell ($R_w=8.6$\,\%).}
\label{fig:partials}
\end{figure}

 All of the PDFs for different configurations of a given composition are very similar (Figure SM-3), and differences are far less significant than termination artifacts introduced in the Fourier transform of $S(Q)$.  A comparison of fits at short $r$ to the PDF of Mn$_{1.23}$Sb for the crystallographically-constrained structure (unit-cell model) versus a representative Mn$_{48}$(Mn$_i$)$_{11}$Sb$_{48}$ supercell is shown in Figure \ref{fig:partials}a. DFT relaxation provides a clear improvement in the description of $G(r)$, reflected in the marked reduction of $R_w$; for fits over the range 2 to 6\,\AA\ the average $R$-factor for all 10 configurations is 13(1)\,\%, whereas $R_w=24$\,\% for the crystallographic structural model. Statistical improvement with the Mn$_{48}$(Mn$_i$)$_{7}$Sb$_{48}$ DFT relaxed structures is also found for fits against the Mn$_{1.15}$Sb PDF data (not shown), with $R_w=13(1)$\,\% versus $R_w=16$\,\%. On the other hand, relaxed Mn$_{48}$(Mn$_i$)$_{4}$Sb$_{48}$ configurations are marginally worse than the crystallographic structure, statistically, in fits against the Mn$_{1.08}$Sb data ($R_w=17.1(4)$\,\% vs. $R_w=15.1$\,\%), but this most likely reflects limited sensitivity to Mn$_i$ correlations for this composition. Indeed, refinement of a Mn$_{1.00}$Sb unit-cell model against the Mn$_{1.08}$Sb data provides a better statistical fit ($R_w=13.5$\,\%) than such a model with partially occupied Mn$_i$.
 
Select atom-pair partials for the crystallographic Mn$_{1.23}$Sb structure and relaxed Mn$_{48}$(Mn$_i$)$_{11}$Sb$_{48}$ supercell are shown in Figure \ref{fig:partials}b. The only pronounced differences occur for Mn$_i$--Sb and Sb--Sb pair correlations, with Mn$_{\mathrm{Mn}}$--Sb and Mn$_{\mathrm{Mn}}$--Mn$_{\mathrm{Mn}}$ partials being notably very similar for the relaxed and unit cell models. As anticipated, equatorial Mn$_i$--Sb correlations are clearly shifted to longer $r$ in the DFT-relaxed model. Differences in the Sb--Sb correlations of the models indicates that the antimony sublattice experiences the largest disruption upon incorporation of Mn$_i$.
 PDF partials for Mn$_i$--Sb and Sb--Sb correlations obtained from reverse Monte Carlo simulation (bottom of Figure \ref{fig:partials}b) agree well with those obtained by DFT, lending additional support to our approach.

 \begin{figure}[!ht]
\centering \includegraphics[width=3.3in]{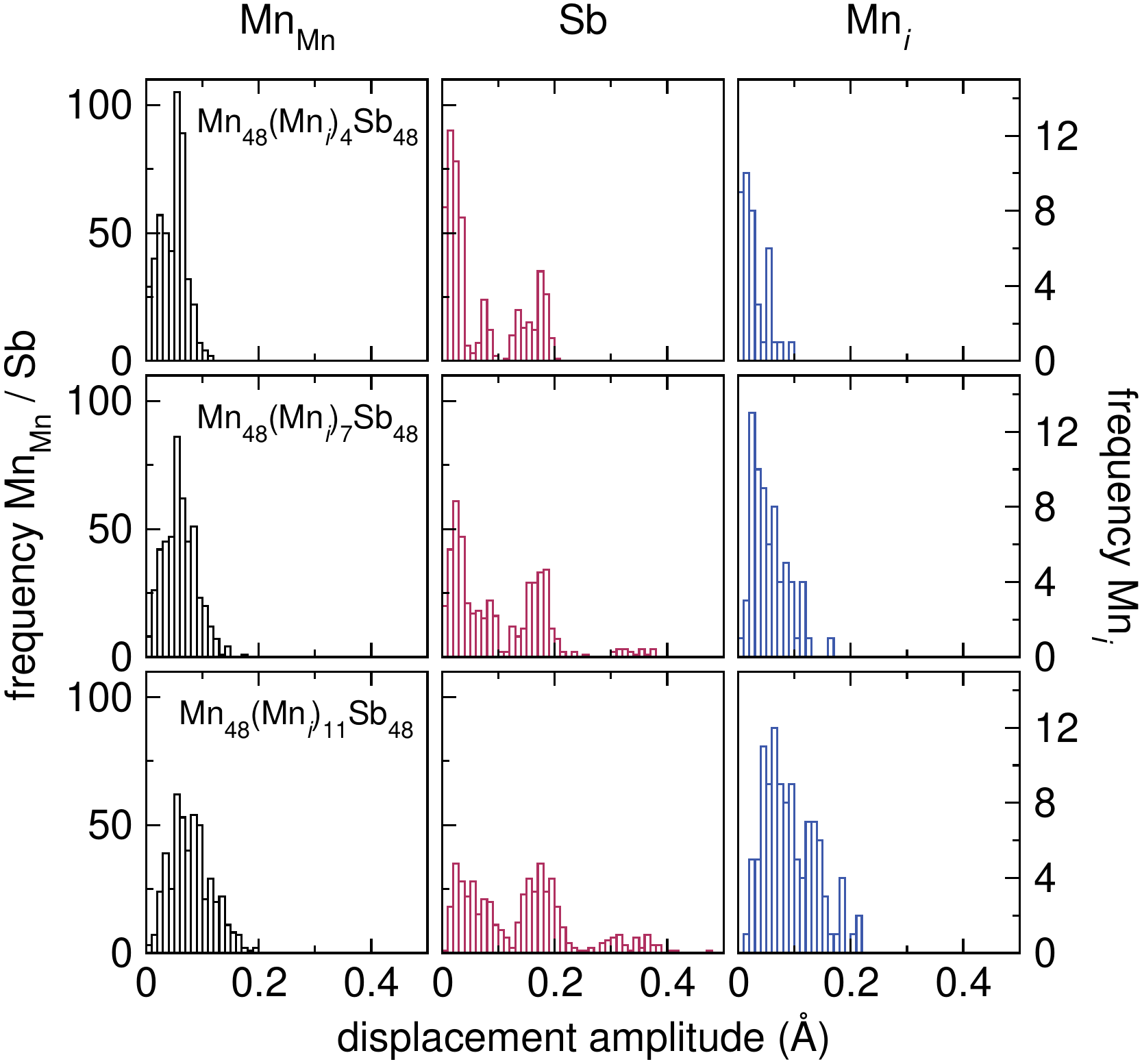}
\caption{Histograms showing the amplitude of atom displacements away from the crystallographic structural positions upon DFT relaxation, compiled for all 10 configurations of each DFT supercell composition. Top row: Mn$_{48}$(Mn$_i$)$_{4}$Sb$_{48}$; middle row: Mn$_{48}$(Mn$_i$)$_{7}$Sb$_{48}$; bottom row: Mn$_{48}$(Mn$_i$)$_{11}$Sb$_{48}$. 
}
\label{fig:displace}
\end{figure}

Histograms displaying the amplitudes of atom displacements away from their crystallographic (average) structural positions upon DFT relaxation are shown in Figure \ref{fig:displace}, arranged by site, and combined for all 10 supercell configurations of each composition. The Mn$_{\mathrm{Mn}}$ and Mn$_i$ sites show unimodal distributions extending to roughly 0.1, 0.15, and 0.2\,\AA, respectively, from lowest to highest Mn$_i$ content. The approximately bimodal Sb histograms show displacements about twice as large for a given composition, with a third region of lower frequency but larger amplitude displacements observed for the Mn$_{48}$(Mn$_i$)$_{7}$Sb$_{48}$  and Mn$_{48}$(Mn$_i$)$_{11}$Sb$_{48}$ configurations associated with Sb that have more than one nearest-neighbor Mn$_i$.  All of the larger amplitude Sb displacements ($\gtrsim0.1$\,\AA) correspond to equatorial Mn$_i$--Sb distances, whereas the displacement of axial Sb atoms is typically $\le 0.1$\,\AA. Despite the large degree of relaxation displayed by Sb atoms, the local distortions necessary to accommodate Mn$_i$ induce very minimal changes in the Mn$_{\mathrm{Mn}}$--Sb and Mn$_{\mathrm{Mn}}$--Mn$_{\mathrm{Mn}}$ bond distance distributions (Figure \ref{fig:partials}b), which are preserved because of cooperative distortions by Mn$_{\mathrm{Mn}}$. This suggests that relaxation should have a much more pronounced effect on the electronic environment of Mn$_i$ than on either Mn$_{\mathrm{Mn}}$ or Sb.

 \begin{figure*}[!ht]
\centering \includegraphics[width=6in]{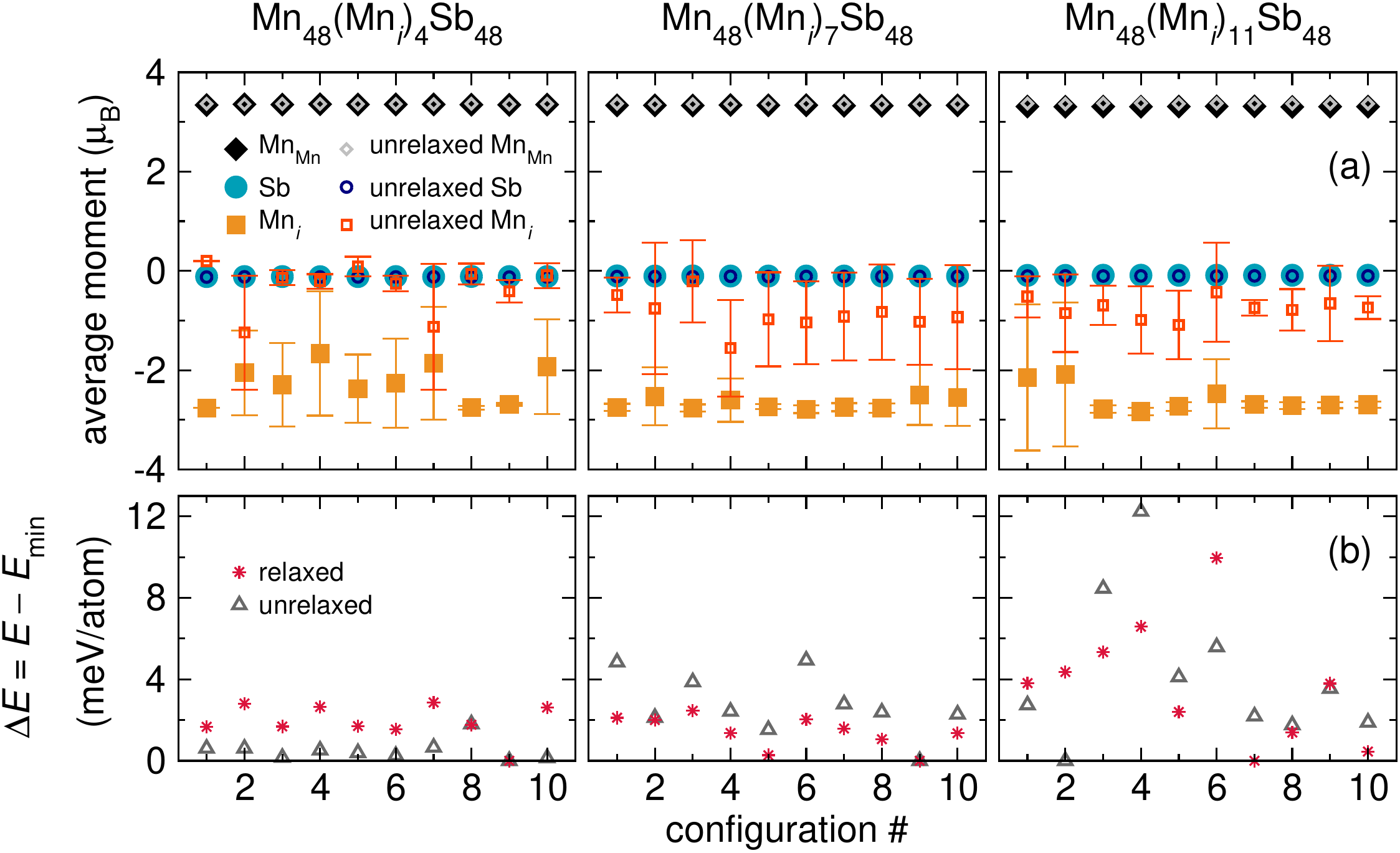}
\caption{(a) Average moment, by site, for each of the relaxed (filled symbols) and unrelaxed (open symbols) configurations among the three compositions examined by DFT. Error bars, representing one standard deviation of the average moments, are much smaller than the symbols for Mn$_{\mathrm{Mn}}$ and Sb. Note that the average moment on Mn$_{\mathrm{Mn}}$ and Sb are quite insensitive to structural relaxation, whereas the magnitude of the moment on Mn$_i$ becomes significantly larger when accounting for relaxation of and around the interstitials.
 (b) Energy of the relaxed (unrelaxed) configurations relative to the most stable relaxed (unrelaxed) configuration for a given composition.
 The energy stabilization of structural relaxation is about 20\,meV\,atom$^{-1}$ for  Mn$_{48}$(Mn$_i$)$_{4}$Sb$_{48}$, 30\,meV\,atom$^{-1}$ for  Mn$_{48}$(Mn$_i$)$_{7}$Sb$_{48}$, and 40\,meV\,atom$^{-1}$ for Mn$_{48}$(Mn$_i$)$_{11}$Sb$_{48}$.
}
\label{fig:moments}
\end{figure*}

The average calculated moments of the Mn$_{\mathrm{Mn}}$, Sb, and Mn$_i$ sites in the DFT relaxed supercells, along with the corresponding moments of unrelaxed supercells, are shown in Figure \ref{fig:moments}a. All 10 configurations are shown for each composition, but there is no relation between configurations of the same number for different compositions. It is immediately apparent that the moments on Mn$_{\mathrm{Mn}}$ and Sb are insensitive and effectively invariant with respect to the different compositions, the different configurations, and also to structural relaxation. For the Mn$_{\mathrm{Mn}}$ and Sb sites, the standard deviations of the average moments are much smaller than the symbols. The average moment on Mn$_i$, however, is consistently of smaller magnitude in unrelaxed configurations than the average Mn$_i$ moment in relaxed supercells, but it is generally opposite in sign to that of Mn$_{\mathrm{Mn}}$. This is reflected in a smaller net magnetization for relaxed structures (Figure \ref{fig:average}b).  For many of the supercells, including both relaxed and unrelaxed structures, there is considerable variability in the individual moments of Mn$_i$ sites, indicated by the error bars which show the standard deviation of the averages.  The individual moments for each Mn$_i$ site in each configuration are shown in Figure SM-4 of the supplemental material. The configuration coordinates are also supplied therein.

Types of variability in the Mn$_i$ moments found in individual configurations include the following: (a) sites in relaxed configurations that remain antiparallel but with decreased magnitude of the moment, (b) sites in relaxed configurations with appreciable magnitude that display parallel alignment with respect to Mn$_{\mathrm{Mn}}$, (c) sites in unrelaxed configurations that are antiparallel and of significantly greater magnitude than the average moment, and (d) sites in unrelaxed configurations that have sizable moments and parallel alignment with Mn$_{\mathrm{Mn}}$. What is surprising about this result is that variability in the Mn$_i$ moments does not necessarily appear to be associated with an energetic penalty. The energies of the relaxed configurations relative to the lowest energy (most stable) relaxed configuration for a given composition are shown in Figure \ref{fig:moments}b; the same comparison is also shown for unrelaxed configurations (relative to the most stable unrelaxed configuration). For example, configurations \#9 are (coincidentally) the lowest energy structures for both Mn$_{48}$(Mn$_i$)$_{4}$Sb$_{48}$ and Mn$_{48}$(Mn$_i$)$_{7}$Sb$_{48}$, including both the unrelaxed and relaxed supercells. Configuration 9 of Mn$_{48}$(Mn$_i$)$_{4}$Sb$_{48}$ is seemingly well-behaved, with large antiparallel moments in the relaxed structure, and small antiparallel moments in the unrelaxed structure. In configuration 9 for Mn$_{48}$(Mn$_i$)$_{7}$Sb$_{48}$, on the other hand, one of the Mn$_i$ sites in the relaxed structure has a significantly smaller (still antiparallel) moment, and two of the sites in the unrelaxed structure have notably larger moments than the average.

Similarly, there are other configurations for each composition that contain substantial variability in the individual Mn$_i$ moments but are lower energy structures than configurations with far less variability. In the case of the relaxed supercells, decreased moments are associated with shorter Mn$_i$--Mn$_{\mathrm{Mn}}$ and Mn$_i$--Sb distances (Figure SM-5). This is consistent with the general observation that the Mn$_i$ moments are smaller in unrelaxed supercells. It does not explain why some of the individual moments are large in the unrelaxed structures; however, it is conceivable that the observed variations result from the imposed periodicities. Nevertheless, the general trends in the Mn$_{\mathrm{Mn}}$, Sb, and Mn$_i$ moments should not be affected significantly.

\subsection{Diffuse scattering and correlated disorder of Mn$_i$}

We now return to the question of the diffuse magnetic scattering observed at the nominally systematically absent (001) position in the recent report by Taylor and co-workers (reference \onlinecite{Taylor:2015}). The signal was observed at the lowest energies accessible in the inelastic experiment ($\sim4$\,meV), but the presence or absence of an associated elastic diffuse signal could not be determined.\cite{Taylor:2015} As noted by the authors, the observation of scattering intensity at this location -- regardless of whether it is elastic or inelastic -- implies symmetry breaking between the upper and lower portions of the unit cell along the $c$ axis. Such breaking could be induced if some Mn$_{\mathrm{Mn}}$ are antiferromagnetically aligned along $c$ (which would double the magnetic unit cell), but both our work and theirs indicates that this is highly improbable. An alternative situation that would break the symmetry is the local population of Mn$_i$ in only one of the two interstitial voids of a unit cell, which removes the $l=2n$ reflection condition. This latter possibility, where interstitial atoms do not necessarily occur in pairs in a given unit cell, was introduced earlier in the discussion of Rietveld refinements (Section III A).

 \begin{figure}[!ht]
\centering \includegraphics[width=3.3in]{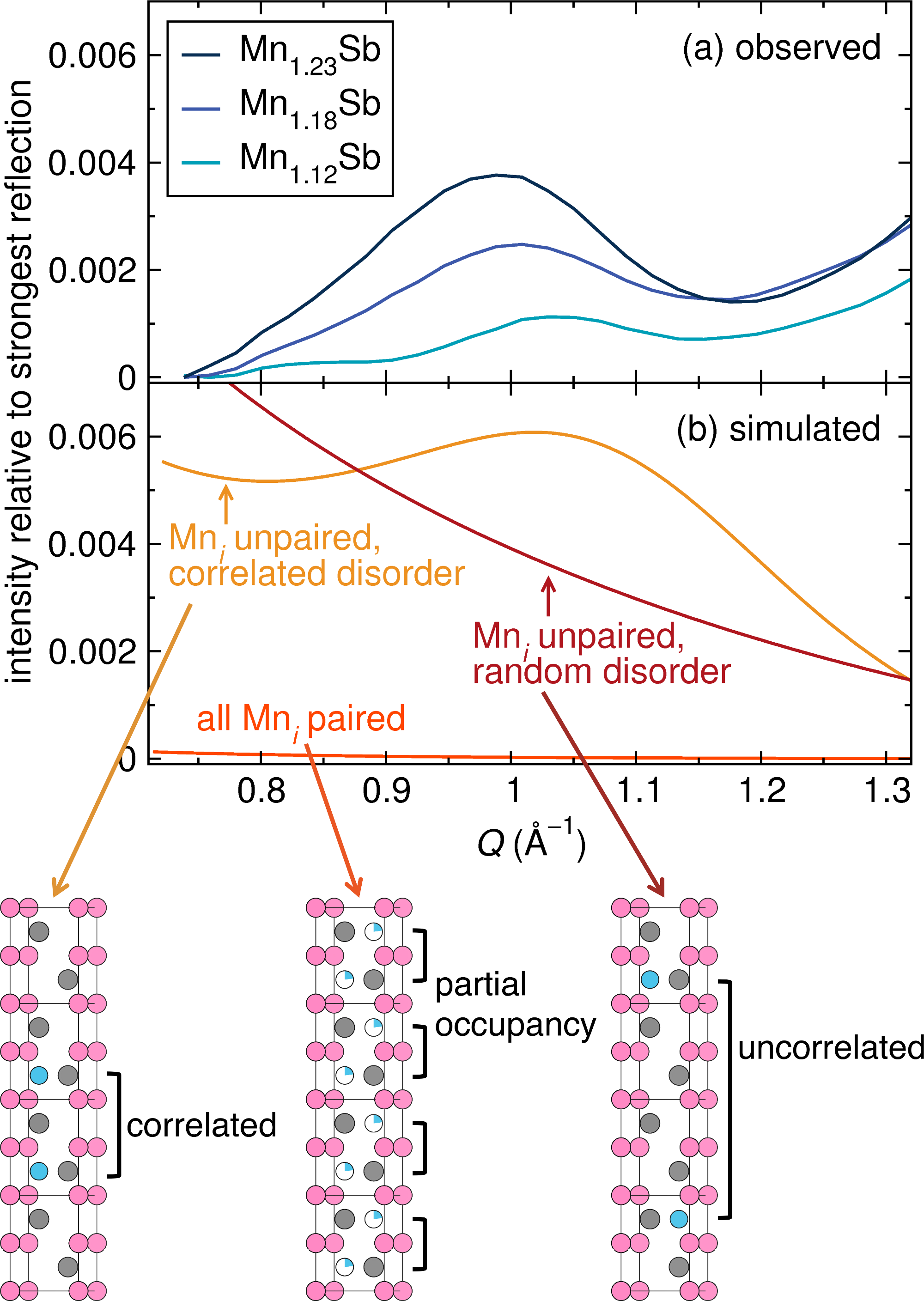}
\caption{(a) Diffuse x-ray scattering features observed in the vicinity of the (001) reflection in Mn$_{1+\delta}$Sb samples ($\delta=0.12$, $0.18$, and $0.23$); container scattering has been subtracted from the data, which was collected at $\sim$\,$86$\,keV at a sample-to-detector distance of $\sim$\,$19$\,cm. (b) Simulated diffraction patterns of Mn$_{1.23}$Sb structures built from an infinite number of infinitely-wide discrete layers. When layers with a singly occupied Mn$_i$ site per unit cell are preferentially stacked adjacent to layers with Mn$_i$ in the same site, the correlated disorder gives rise to a diffuse feature that is consistent with the experimentally observed scattering in regards to both position (in $Q$) and relative intensity. Cartoon models of the different structures giving rise to the simulated scattering are shown as four layer fragments; Mn$_{\mathrm{Mn}}$ (pink), Sb (gray), Mn$_i$ (blue).
}
\label{fig:diffuse}
\end{figure}

Close examination of our x-ray scattering data reveals a weak diffuse component on top of the systematically absent (001) reflection that is suggestive of correlated structural disorder. In Figure \ref{fig:diffuse}a we present the low-$Q$ region of x-ray diffraction patterns for three Mn$_{1+\delta}$Sb samples ($\delta=0.12$, $0.18$, and $0.23$), from which the container scattering has been subtracted. Scattering intensities have been normalized to the most intense peak in the x-ray diffraction pattern of Mn$_{1+\delta}$Sb [\textit{i.e.}\ the (101) reflection]. The intensity of the diffuse signal increases with $\delta$, and the maxima of the diffuse features are all observed at lower $Q$ (larger $d$-spacing) than the center of the (001) reflection on the basis of the $c$ lattice constant. 
A shift of diffuse scattering maxima away from the associated Bragg positions is commonly observed in situations of local structural dilation or contraction.\cite{Dederichs:1971}

A series of simulations are presented in Figure \ref{fig:diffuse}b that offer support to the hypothesis that the diffuse signal at low $Q$ arises from correlated structural disorder associated with ``unpaired'' Mn$_i$, that is, occupation of just one of the interstitial sites in a unit cell. Diffraction patterns were generated with the DIFFaX software\cite{TREACY:1991} -- which explicitly computes the incoherent intensity contribution to a given $hkl$ reflection -- for structures comprising random and correlated stacking variations of discrete layers. All simulations were conducted on the basis of the Mn$_{1.23}$Sb stoichiometry and cell dimensions.

As a control DIFFaX simulation, a single-layer Mn$_{1+\delta}$Sb model containing partially occupied Mn$_i$ in both of the interstitial sites was considered. This model is designated ``all Mn$_i$ paired'' in Figure \ref{fig:diffuse}b, and it results in no noticeable scattering intensity at the (001) position. Models containing unpaired Mn$_i$ were constructed with three distinct layer types: (1) a layer without any interstitial atoms, (2) a layer with a fully occupied interstitial atom in the lower interstitial void, and (3) a layer with a fully occupied interstitial atom in the upper interstitial void. Two types of disorder were modeled for structures incorporating layers with singly occupied interstitial voids, namely random disorder and correlated disorder, as labeled in Figure \ref{fig:diffuse}. For the model containing layers with unpaired Mn$_i$ stacked in a completely random assortment, no resolved diffuse intensity is generated at the (001) reflection. The rise in intensity at low $Q$ is the result of the diffuse scattering from completely incoherent contributions, which emanate from (000). On the other hand, for an unpaired Mn$_i$ model in which there is preferential stacking of the layers -- correlated disorder -- a well resolved diffuse signal near the (001) is generated. More specifically, the preferential stacking modeled here corresponds to an increased probability that a layer containing a single Mn$_i$ in one of the interstitial voids will neighbor layers that also contain Mn$_i$ in the same site. Although this simulation represents a rather coarse example of one type of correlated disorder that could be present in Mn$_{1+\delta}$Sb, it offers strong support for the presence of some degree of clustering associated with the interstitial atoms.  As noted for the experimentally observed diffuse signal, the maximum of the diffuse component in the simulated pattern is also shifted to lower $Q$ than the center of the (001) reflection. A single-crystal x-ray diffuse scattering study would be better suited to elucidating the origin of this observation.

While these observations and simulations soundly establish the existence of short-range structural order, the specific nature of clustering in Mn$_{1+\delta}$Sb remains an open question. 
Consideration of the many plausible local arrangements is beyond the scope of the present work; it is easily conceivable that a variety of short range-ordered motifs coexist. It is important to stress that although the presence of some unpaired Mn$_i$ is a requisite for generating nuclear scattering intensity near the (001) reflection, this does not imply the absence of local regions with pairs of interstitials. Indeed, this type of diversity is represented in the DFT supercells. A comparison of the relative supercell stabilities does not suggest an obvious penalty for interstitial proximity (pairing), but because of their relatively small sizes and construction by stochastic population of the interstitial sites it would be conjecture to suggest an energetic preference for any particular type of clustering.

In summary, the observation of a diffuse signal consistent with the symmetry-forbidden (001) reflection suggests that the diffuse magnetic component observed by Taylor \textit{et al.}\ by inelastic neutron scattering is rooted in short-range structural order.  This explanation is further supported by our density functional studies, which challenge previous notions\cite{YAMAGUCHI:1976,Watanabe:1980,YAMAGUCHI:1980,COEHOORN:1985}
 that Mn$_i$ induces a significant reduction in neighboring Mn$_{\mathrm{Mn}}$ moments.

\section{Conclusions}

Analysis of x-ray total scattering data collected on polycrystalline powders of Mn$_{1+\delta}$Sb ($0.03 \le \delta \le 0.23$) reveals signatures of structural relaxation associated with the accommodation of interstitial atoms.  In particular, the trigonal bipyramidal site of Mn$_i$ is better described by longer equatorial Mn$_i$--Sb distances than reflected by the crystallographic unit cell.  Density functional relaxation of large supercells for compositions close to $\delta = 0.08$, $0.15$, and $0.23$, provides excellent structural descriptions of the short-range disorder induced by Mn$_i$, evidenced by  improved fits to the experimental PDFs. Relaxation has a pronounced effect on the calculated moment of Mn$_i$, which aligns antiparallel to Mn$_{\mathrm{Mn}}$, but our results suggest that Mn$_i$ has effectively no influence on the moments of Mn$_{\mathrm{Mn}}$ in proximity to interstitials. The observation of a diffuse  signal near the (001) reflection position, which is systematically absent when Mn$_i$ is distributed completely randomly, indicates that some degree of {\em correlated} disorder exists in Mn$_{1+\delta}$Sb. \\

\section*{acknowledgments}
We gratefully acknowledge Dr.\ Kate A.\ Ross for useful discussions, Dr.\ Olaf J.\ Borkiewicz for assistance collecting total scattering data, and the instrument scientists at 11-BM for powder diffraction data acquisition. Use of the Advanced Photon Source at Argonne National Laboratory was supported by the U. S. Department of Energy, Office of Science, Office of Basic Energy Sciences, under Contract No. DE-AC02-06CH11357. This research utilized the CSU ISTeC Cray HPC System supported by NSF Grant CNS-0923386.


%

\end{document}